# Specular Reflections from Artificial Surfaces as Technosignature

Bhavesh Jaiswal[1,2]

**Abstract**

Direct imaging of exoplanets will allow for observation of a planet in reflected light. Such a scenario may eventually allow for the possibility to scan a planetary surface for the presence of artificial structures made by alien civilizations. Detectability of planetary-scale structures, called megastructures, has been previously explored. In this work, we show that it is possible to detect structures of much smaller scale on exoplanetary surfaces by searching for the specular reflection of host starlight from the corresponding structures. As the planet rotates, these reflections can manifest as an optical transient riding atop the rotational light curve of the planet. Due to the directional nature of specular reflection, the reflected signal is very strong, and it is comparable to the planetary flux for surfaces that cover only a few parts per million of the total planetary surface area. By tracking the planet around its orbit, it should be possible to scan the planetary surface for any such structures that cover a size larger than a few parts per million of planetary surface. The proposed method will aid in the search for extraterrestrial intelligence in the era of direct imaging of exoplanets. Key Words: Alien civilization—Specular reflection—Exoplanet.

F<small>UTURE EFFORTS TO SEARCH</small> for and characterize nearby exoplanets will be based on directly detecting the reflected light from the planet by blocking the light from the host star. This strategy is being considered for space missions (Lacy *et al.,* 2019; https://www.jpl.nasa.gov/habex; https://asd.gsfc.nasa.gov/luvoir) as well as large ground-based telescopes (Chauvin, 2018; Fuji *et al.,* 2018). In the era of discovering planets via direct imaging, it is anticipated that investigators will analyze planets for the presence of large-scale features such as continents, oceans, and vegetation via photometry and spectroscopy (Cowan *et al.,* 2009; Zugger *et al.,* 2010; Livengood *et al.,* 2011). It may be possible as well to detect signatures of planetary-scale artificial megastructures due to the effect of their albedo on reflected light (Lingam and Loeb, 2017; Berdyugina and Kuhn, 2019). Such signatures would be termed ''technosignatures.'' Wright *et al.* (2022) argued that the abundance of technosignatures could be much higher than previously anticipated and that they can be long-lived and highly detectable. In this paper, we consider specular reflections from artificial and smooth surfaces as a viable technosignature.

Artificial constructions on a planetary surface could be created by an alien civilization for various purposes. They could serve as a way of harnessing energy (such as solar panels) or be a part of a bigger engineering structure or a space for cohabitation (such as cities). They could constitute other types of constructions as well that lie beyond our present knowledge and understanding. Such constructions could be in the form of a single large structure or an agglomeration of several smaller structures. We consider a case where these structures have a flat and smooth surface capable of reflecting the host starlight in a specular fashion. Specular reflection of starlight can be about a million times stronger than the Lambertian reflection from a planetary surface. This is because the reflected light preserves the original divergence in the case of specular reflection as opposed to Lambertian reflection where the reflected light is spread in $2\pi$ steradians. The reflected signal is proportional to the reflectivity ($r$) of the structure, which we consider spanning from very low to very high values. Glass ($r=0.05$) and aluminum ($r=0.9$) (Hecht, 2017) are considered as two examples of very low and very high reflectivity materials, respectively. The surface can be either made or covered with these structures. For example, transparent glass can be used as a cover for solar panels or as a roof in an artificial greenhouse. High-reflectivity material such as aluminum sheets can be used to reflect the stellar energy back to space to control ambient temperatures.

---

[1]Space Astronomy Group, U R Rao Satellite Centre, Bangalore, India.
[2]Department of Physics, Indian Institute of Science, Bangalore, India.

Figure 1 (left) shows the geometry of the direct reflection from such a large structure on a planet. For a general case, the structure is shown to be fragmented ($A_1$, $A_2$, $A_3$). While calculating the reflections from the flat surfaces of large structures, the curvature introduced by the planetary surface should also be considered. Consider the reflection from the structure $A_1$. The structure $A_1$ is merely steering the beam of the star in the direction of the observer. The distance of the star from the planet is infinitely smaller than the distance of the observer from the planet. Due to this, only plane-parallel rays from $A_1$ toward the observer are considered. These rays, if propagated backward, would intercept an area on the star equal to the area of $A_1$ projected in the direction of the star. Due to the curvature of the planet, different structures will reflect light from different portions of the stellar disc.

To calculate flux from the structures and the planet, the stellar intensity is considered as $I$, the stellar radius as $R_S$, the planetary radius as $R_P$, the planet-star separation as $a$, and the distance to this planetary system as $d$. The intensity reflected by the specular reflection from the artificial structure is then $r*I$. The reflected flux would then be dependent on the solid angle of the structure. The "maximum" solid angle (when $\Theta_A \approx 2R_S/a$) of the structure at the observer in the full-phase scenario is given by

$$\Omega_S = \pi \left(\frac{R_P R_S}{ad}\right)^2 \quad (1)$$

Thus, the specularly reflected flux of the artificial structure at distance $d$ is given by

$$F_S = Ir\pi \left(\frac{R_P R_S}{ad}\right)^2 \quad (2)$$

Now, consider the planetary flux ($F_P$) for a full phase scenario for a planet having a geometrical albedo of $A_g$. It is given by (Seager, 2010)

$$F_P = IA_g\pi \left(\frac{R_P R_S}{ad}\right)^2 \quad (3)$$

Hence, $F_S$ can be written as

$$F_S = \frac{r}{A_g} F_P \quad (4)$$

This shows that the reflected flux from a specularly reflecting structure can be comparable or even larger than the planetary flux ($A_g \sim 0.3$ for Earth) despite its much smaller area. The total flux from the planet ($F$) is given by $F = F_S + F_P$. In Fig. 1, we show this variation for a rotating Earth-like planet for a 90° view angle and for the two values of structure reflectivity, that is, glass ($r = 0.05$) and aluminum ($r = 0.9$). The planetary flux ($F_P$) has been normalized to 1.

Consider the Sun-Earth system where the angular extent of the Sun is 0.52° from Earth. As shown in the figure, the angle of incidence of starlight on $A_1$ is equal to $\Theta_1$. Due to the reflection geometry, the expected range in the variation of $\Theta_1$, $\Theta_2$, and $\Theta_3$ is 0.52°/2 (= 0.26°). This is also the same as the extent of $\Theta_A$ (= 0.26°). The extent of $\Theta_A$ defines the size of the total structure area on the surface of the planet. It is noteworthy that the total area of the specularly reflecting structure must fit into an area of the planet that can reflect the starlight in the direction of the observer. This area is equal to a circle with diameter $\sim R_P \times \Theta_A$ on the surface of the planet (Fig. 1). This is the maximum area of structure that can cause specular reflection at a time. For a planet like Earth, this would correspond to a total area of about 2800 km$^2$, which is only 5.4 ppm of the total surface area of the planet (about the size of a modern-day city).

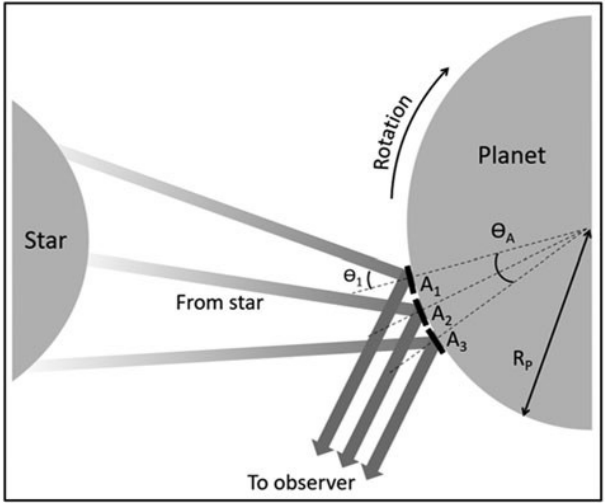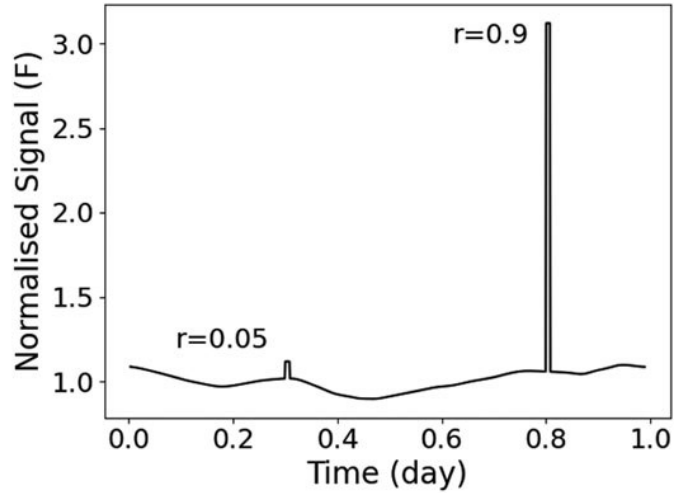

**FIG. 1.** (**Left**) The geometry of specular reflection from the structures present on the surface of the planet. The artificial structures are drawn on the surface (marked as $A_1$, $A_2$, and $A_3$). The star has a finite size as seen from the planet; hence the incident rays will have a variation in incident angles. The reflected rays, however, are parallel in the direction of the observer as the distance of the planet from the observer is very large. Only those rays that are reflected in the direction of the observer are shown here. (**Right**) The normalized light curve of rotating Earth is shown along with the specular reflection from the artificial structure. The Earth light curve is taken from Livengood *et al.* (2011). The *x*-axis corresponds to the planetary rotation in fraction of days, with 1 corresponding to a complete rotation. The reflection peaks are seen as sharp, narrow, and bright spikes in the light curve. The spike observed at 0.3 days corresponds to low reflectivity ($r = 0.05$), and the one at 0.8 days corresponds to high reflectivity structure ($r = 0.9$). The total surface area of both structures is spread between ±1.5° longitude and ±0.15° latitude.

Any structures present outside this area will not appear to be illuminated to the observer as the light from no part of the star can be reflected in the direction of the observer. Structures present outside this area will require a small tilt in order to reflect the light in the direction of the observer. As shown in Fig. 1, variations in the light curve of the planet are due to large continents, oceans, and cloud systems, which appear and disappear as the planet rotates. The specular reflection peaks from glass and aluminum surfaces are marked in the light curve. The specular reflection from glass is faint due to very low reflectivity, but it is still noticeable. The reflection from a metal like aluminum would be seen to be several times brighter than the planet. Here, the entire structure ($A_1$, $A_2$, and $A_3$) is assumed to lie in between $\pm 1.5°$ longitude and $\pm 0.15°$ latitude (encompassing a total area of about 20 ppm of the entire planetary surface). Although for bright reflection spikes, only a small area of the structure is sufficient (5 ppm; corresponding to $\Theta_A$), but such spikes would be very sharp and could be missed on a planet that has a fast rotation. A larger longitudinal span of a structure would prolong the duration of the bright spikes as the planet rotates, though there is no further increase in the brightness due to the increased area of the structure. To increase the chances of detection, an alien civilization could create a reflective structure with a sufficient longitudinal width to produce the brightness of sufficient duration and a latitudinal extent from pole to pole to ensure the detectability of the reflection in all directions. It is worth mentioning that the specular reflection peaks in the light curve are expected to have a large degree of linear polarization due to the process of reflection, which can also serve to confirm the nature of reflection if such reflections are ever observed from a planet.

Due to the curvature of a planetary surface and the narrow angular size of the star as seen from the planet, the specular reflections of starlight on the structure occur at a very small area on the planet. As the planet rotates, it is necessary that the reflecting structure fall within a narrow range of permissible latitudes from where the specular reflection can be directed toward the observer. For example, for the light curve in Fig. 1 (right), the structure is assumed to be located on the equator within $\pm 0.15°$ latitude. If the structure is above or below these latitudes, the reflection can be missed. However, it is possible to scan different latitudes by utilizing the tilt in the observation geometry. This can be caused by a tilt either in the planet's rotation axis or in the orbital plane. For example, if the planet's axis of rotation has a tilt (like Earth's tilt of 23.5°) with respect to the star-planet plane, it would be possible to scan different latitudes at different locations of the planet in its orbit. Geometry of such scanning is illustrated in Fig. 2 for an orbit that is near parallel to the observing direction. In the case of the Sun-Earth system seen from the direction of vernal equinox, it would be possible to scan 23.5° of latitudes above and below the equator. In a similar way, scanning would be possible for a planet that has no tilt in the rotation axis but does have an orbital plane that is tilted with respect to the direction of observation.

Exoplanets that have axial/orbital tilts and cloud-free atmospheres would be ideal targets for scanning the planetary surface for such specular reflections. For the reflection to be detectable, the telescope integration time would have to be similar to, or less than, the width of the specular reflection peak. For this reason, the structures present on planets that have a slow rotation speed or structures that have a large longitudinal spread would have better chances of detection. The structures considered here for calculations are assumed to have a perfect orientation, that is, locally parallel to the planetary surface. However, small-level mutual misalignments do not change the results significantly. The acceptable range of mutual misalignment can be of the same size as the angular size of the stellar disc from the planet if the magnitude of the specular reflection needs to remain unchanged.

Future efforts to directly image planets in reflected light are being strongly pursued for both space telescopes and for extremely large ground-based telescopes. The space missions largely focus on Sun-like stars where the planet-star contrast for an Earth-like planet in the habitable zone is expected to be $\sim 1 \times 10^{-10}$. The plans for ground-based telescopes mainly pursue M-dwarf stars, for which the contrast requirements of $\sim 1 \times 10^{-8}$ can be sufficient (Chauvin, 2018; Fuji *et al.*, 2018). Calculations of specular reflection from artificial structures in the M-dwarf planetary systems also yield similar results, and the reflected signal is comparable to the planetary flux.

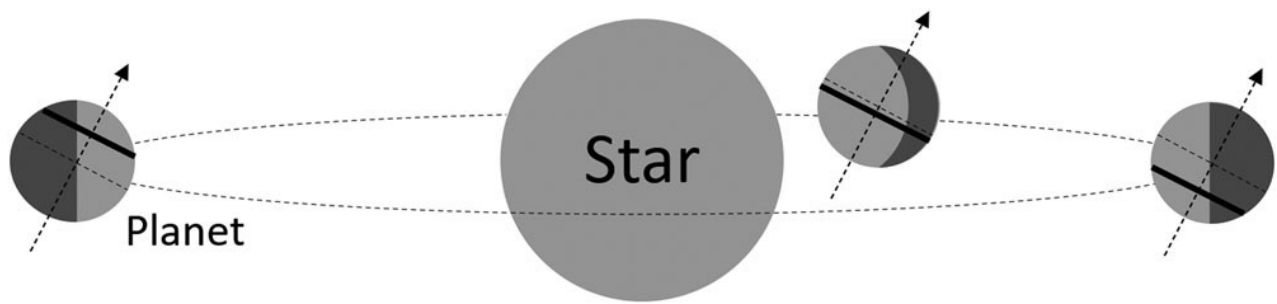

**FIG. 2.** Location of the planet at different positions in its orbit allows for scanning different latitudes for the presence of specularly reflecting artificial structures. The planet's rotation axis is shown to be tilted with respect to the orbital plane. The axis of rotation is marked with an arrow, and the equator is marked with a dashed line. At each location in the orbit of the planet, the latitude, which can be scanned for specular reflections as the planet rotates, is marked as a solid line. This latitude allows the specular reflection in the direction of the observer, which is perpendicular to this page.

### Acknowledgment

The author thanks the reviewer for many suggestions which have helped improve the quality of this manuscript.

### References

Berdyugina SV, Kuhn JR. Surface imaging of Proxima b and other exoplanets: Albedo maps, biosignatures, and technosignatures. *Astron J* 2019;158(6); doi: 10.3847/1538-3881/ab2df3.

Chauvin G. Direct imaging of exoplanets at the era of the extremely large telescopes. *arXiv e-prints,* 2018; arXiv:1810.02031.

Cowan NB, Agol E, Meadows VS, *et al.* Alien maps of an ocean-bearing world. *Astrophys J* 2009;700(2):915–923; doi: 10.1088/0004-637X/700/2/915.

Fuji Y, Angerhausen D, Deitrick R, *et al.* Exoplanet biosignatures: Observational prospects. *Astrobiology* 2018;18(6): 739–783; doi: 10.1089/ast.2017.1733.

Hecht B. *Optics*. Pearson Education: Boston; 2017.

Lacy B, Shlivko D, Burrows A. Characterization of exoplanet atmospheres with the optical coronagraph on WFIRST. *Astron J* 2019;157(3); doi: 10.3847/1538-3881/ab0415.

Lingam M, Loeb A. Natural and artificial spectral edges in exoplanets. *Mon Not R Astron Soc Lett* 2017;470(1):L82–L86; doi: 10.1093/mnrasl/slx084.

Livengood TA, Deming LD, A'hearn MF, *et al.* Properties of an Earth-like planet orbiting a Sun-like star: Earth observed by the EPOXI mission. *Astrobiology* 2011;11(9):907–930; doi: 10.1089/ast.2011.0614.

Seager S. *Exoplanet Atmospheres: Physical Processes*. Princeton University Press: Princeton, NJ; 2010.

Wright JT, Haqq-Misra J, Frank A, *et al.* The case for technosignatures: Why they may be abundant, long-lived, highly detectable, and unambiguous. *Astrophys J Lett* 2022;927(2): L30; doi: 10.3847/2041-8213/ac5824.

Zugger ME, Kasting JF, Williams DM, *et al.* Light scattering from exoplanet oceans and atmospheres. *Astrophys J* 2010;723(2):1168–1179; doi: 10.1088/0004-637X/723/2/1168.

Address correspondence to:
*Bhavesh Jaiswal*
*Space Astronomy Group*
*U R Rao Satellite Centre*
*Bangalore, 560037*
*India*